# Energy-Efficient Cooperative Cognitive Relaying Schemes for Cognitive Radio Networks

Ahmed El Shafie, *Student Member, IEEE,* Tamer Khattab, *Member, IEEE,* Amr El-Keyi, *Member, IEEE*

*Abstract*—We investigate a cognitive radio network in which a primary user (PU) may cooperate with a cognitive radio user (i.e., a secondary user (SU)) for transmissions of its data packets. The PU is assumed to be a buffered node operating in a time-slotted fashion where the time is partitioned into equal-length slots. We develop two schemes which involve cooperation between primary and secondary users. To satisfy certain quality of service (QoS) requirements, users share time slot duration and channel frequency bandwidth. Moreover, the SU may leverage the primary feedback message to further increase both its data rate and satisfy the PU QoS requirements. The proposed cooperative schemes are designed such that the SU data rate is maximized under the constraint that the PU average queueing delay is maintained less than the average queueing delay in case of non-cooperative PU. In addition, the proposed schemes guarantee the stability of the PU queue and maintain the average energy emitted by the SU below a certain value. The proposed schemes also provide more robust and potentially continuous service for SUs compared to the conventional practice in cognitive networks where SUs transmit in the spectrum holes and silence sessions of the PUs. We include primary source burstiness, sensing errors, and feedback decoding errors to the analysis of our proposed cooperative schemes. The optimization problems are solved offline and require a simple 2-dimensional grid-based search over the optimization variables. Numerical results show the beneficial gains of the cooperative schemes in terms of SU data rate and PU throughput, average PU queueing delay, and average PU energy savings.

## I. INTRODUCTION

Secondary utilization of the licensed frequency bands can efficiently improve the spectral density of the under-utilized licensed spectrum. Cognitive radio (secondary) users are intelligent devices that use cognitive technologies to adapt with variations, and exploit methodologies of learning and reasoning to dynamically reconfigure their communication parameters [2]–[4]. This allows the secondary users (SUs) to utilize the spectrum whenever it is free to use and with the maximum possible data rates.

Cooperative diversity is a recently emerging technique for wireless communications that has gained wide interest [5]–[8]

Part of this paper was published in the IEEE International Conference on Computing, Networking and Communications (ICNC), 2015 [1].

A. El Shafie is with the University of Texas at Dallas, USA (e-mail: ahmed.elshafie@utdallas.edu).

T. Khattab is with Electrical Engineering, Qatar University, Doha, Qatar (email: tkhattab@ieee.org).

A. El-Keyi is with Wireless Intelligent Networks Center (WINC), Nile University, Giza, Egypt (aelkeyi@nileuniversity.edu.eg).

The work of T. Khattab is supported by Qatar National Research Fund (QNRF) under grant number NPRP 7-923-2-344. The statements made herein are solely the responsibility of the authors.

where multiple channels are used to communicate the same information symbol. Recently, cooperation in cognitive radio networks, referred to as the cooperative cognitive relaying, where the SU helps in relaying some of the undelivered primary user (PU) packets, has got extensive attention [9]–[16]. In particular, the SU functions as a relay node for the PU whenever the PU packet cannot be decoded at its destination. The authors of [9] showed that the maximum achievable rate can be achieved by simultaneous transmissions of PU and SU data signals over the same frequency band. The SU data signals are jointly encoded with PU data signals via dirty-paper coding techniques. Hence, the SUs know perfectly the PU's data. In [10], the authors assumed that the SU decodes-and-forwards the undelivered PU packets during the idle periods of the PU. The SU maximizes its throughput by adjusting its transmit power level.

### A. Related Work

In [12], the authors investigated the scenario of deploying a *dumb* relay node in cognitive radio networks to increase network spectrum efficiency. The relay node aids both the PU and the SU. The proposed scheme is investigated for a network consisting of a pair of PUs and a pair of SUs. In [13], the authors considered a network with one buffered PU and one buffered SU where the SU is allowed to access the channel when the PU's queue is empty. The SU has a relaying queue to store a fraction of the undelivered PU packets controlled through an adjustable admittance factor. A priority of transmission is given to the relayed PU packets over the SU own packets. The SU aims at minimizing its average queueing delay subject to a power budget for the relayed primary packets. In [15], the authors characterized some fundamental issues for a wireless shared channel composed of one PU and one SU. The authors considered a general multi-packet reception model, where concurrent packet transmission could be correctly decoded at receivers with a certain probability that is characterized by the system's parameters (e.g., packet length, data rates, time slot duration, bandwidth, etc.). The PU has unconditional channel access, whereas the SU accesses the channel based on the activity state of the PU, i.e., active or inactive, during a time slot. The spectrum sensing process is impractically assumed to be perfect. The SU is assumed to be capable of relaying the undelivered PU packets as in [13]. If the PU is sensed to be inactive during a time slot, the SU accesses the channel with probability one, and if the PU is active, the SU randomly accesses the channel simultaneously with the PU or attempts to decode the primary packet with

4the complement probability. The maximum stable throughput region of the network is obtained via optimizing over the access probability assigned by the SU during the active periods of the PU.

Releasing portions of primary systems time slot duration and bandwidth for the SUs has been considered in several works, e.g., [11], [14], [17]. In [11], the authors proposed a spectrum leasing scheme in which PUs may lease their owned bandwidth for a fraction of time to SUs based on decode-and-forward (DF) relaying scheme and distributed space-time coding. In [14], the authors proposed a new cooperative cognitive scheme, where the PU releases portion of its bandwidth to the SU. The SU utilizes an amplify-and-forward relaying scheme. It receives the primary data during the first half of the time slot, then forwards the amplified data during the second half of the time slot. In [17], the authors considered an SU equipped with multiple antennas sharing the spectrum with a single-antenna energy-aware PU, where the PU aims at maximizing its mean transmitted packets per joule. The users (SU and PU) split the time slot duration and the total bandwidth to satisfy certain quality of service (QoS) for the PU that cannot be attained without cooperation. Both users maintain data buffers and are assumed to send one data packet per time slot.

*B. Contributions*

Given the need for shorter transmission times and low latency communications [18]–[20], we develop two cooperative cognitive schemes which allow the SU to transmit its data bits simultaneously with the PU under the constraint of short communication times and the presence of practical sensing and feedback cost considerations. Under our proposed schemes, the PU may cooperate with the SU to enhance its QoS, i.e., to enhance its average queueing delay and maintain its queue stability. Hence, cooperation is optional for the PUs. If cooperation is beneficial for the PU, it releases portion of its bandwidth and time slot duration for the SU. In turn, the SU incurs portion of its transmit energy to relay the primary packets. The SU employs a DF relaying scheme. The time slot is divided into several intervals (or time phases) that change according to the adopted cooperative scheme, as will be explained later. In our first proposed cooperative scheme, the SU *blindly* forwards what it receives from the PU even if the primary destination can decode the data packet correctly at the end of the PU transmission phase. On the other hand, in our second proposed scheme, the SU forwards what it receives from the PU if and only if the primary destination could not decode the PU transmission of the primary packet; or if the SU considers the feedback message as a negative-acknowledgement from the primary destination.[1] However, as will be explained later, there is a cost for using the second cooperative scheme which is a reduction in the time available for transmission data of users due to the presence of an additional feedback duration. These practical issues are quantified analytically in this work.

The contributions of this paper are summarized as follows
- We design two cooperative cognitive schemes which involve cooperation between the PUs and the SUs. The two schemes differ in terms of time slot structure and primary feedback mechanism. Both schemes achieve a significant PU energy savings.
- We consider practical assumptions for the cognitive radio network. Precisely, unlike most exiting literature, we consider spectrum sensing errors and primary feedback reception errors at the SU. Moreover, we consider the impact of the time durations spent on spectrum sensing and feedback message transmission on the achievable data rates. In addition, the PU data burstiness is taken into consideration.
- We propose two QoS measures for the PU and include them in the proposed optimization problems as constraints. Specifically, we assume a constraint on the PU average queueing delay and a constraint on the stability of the PU queue. Moreover, we consider a practical energy constraint on the SU average transmit energy. The optimization problems are stated under such constraints.

This paper is organized as follows. In the next section, we introduce the system model adopted in this paper. In Section III, we analyze the PU queue and derive the PU average queueing delay and PU queue stability condition. Our first proposed cooperative scheme is explained in Section V. In Section VI, we describe our second proposed cooperative scheme. The numerical results are shown in Section VIII. We finally conclude the paper in Section IX.

## II. SYSTEM MODEL

We consider a wireless network composed of orthogonal primary channels, where each channel is used by one PU. Each primary transmitter-receiver pair coexists with one secondary transmitter-receiver pair. For simplicity, we focus on one of those orthogonal channels.[2] Each orthogonal channel is composed of one secondary transmitter 's', one primary transmitter 'p', one secondary destination 'sd' and one primary destination 'pd'. The SU is equipped with two antennas: one antenna for transmission data and the other for data reception and spectrum sensing. The PU is equipped with a single antenna. Moreover, the PU has an infinite-length buffer for storing a fixed-length packets. The arrivals at the PU queue are independent and identically distributed (i.i.d.) Bernoulli random variables from one time slot to another with mean $\lambda_{\rm p} \in [0,1]$ packets per time slot. Thus, the probability of a data packet arrival at the PU queue in an arbitrary time slot is $\lambda_{\rm p}$. A list of the key variables is given in Table I.

*A. Channel Model*

We assume an interference wireless channel model, where concurrent transmissions are assumed to be lost data if the received signal-to-noise-plus-interference-ratio (SINR) is less than a predefined threshold, or equivalently, if the instantaneous channel gain is lower than a predefined value.[3] We propose a DF relaying technique, where the SU decodes and then forwards the PU packet. The SU is assumed to be a full-duplex terminal which means that it can receive and transmit at the same time. To avoid the loopback self-interference impairments which can significantly reduce the achievable rates, we assume that the SU cannot transmit and receive over the same frequency band. However, the SU can transmit data over a frequency band and receive over the other.

---

[1]In this paper, the primary feedback channel is assumed to be modeled as an erasure channel model and can be undecodable at the secondary terminal. This will be justified in Section VI.

[2]As argued in the cognitive radio literature, e.g., [9]–[17] and the references therein, the proposed cooperative cognitive scheme and theoretical development presented in this paper can be generalized to cognitive radio networks with more PUs and more SUs.

TABLE I: List of Key Variables.

| Symbol | Description | Symbol | Description |
| --- | --- | --- | --- |
| $\tau_\text{s}$ | Spectrum sensing time duration | $T$ and $W$ | Time slot (coherence time) duration and channel total bandwidth, respectively |
| $\tilde{\mathcal{R}}$ | Average SU data rate | $P_\text{o}$ | Average transmit information power |
| $Q_\text{p}$ | Queue at the PU | $\tau_f$ | Feedback message duration |
| $\mathcal{R}_\text{e}^{(\ell)}$ and $\mathcal{R}_\text{b}^{(\ell)}$ | SU transmission data rate under scheme $\mathcal{P}_\ell$ when the PU queue is empty and nonempty, respectively | $\mu_\text{p,c}^{(\ell)}$ | Average service rate of the PU queue under scheme $\mathcal{P}_\ell$ |
| $\alpha_\text{j,k}$ | Channel gain of the j − k link with mean $\sigma_\text{j,k}^2$ | $P_\text{FA}$ | False alarm probability at the SU |
| $P_\text{MD}$ | Misdetection probability at the SU | $\lambda_\text{p}$ | Average arrival rate at the PU's queue |
| $D_\text{p,c}^{(\ell)}$ | Average queueing delay at the PU queue under scheme $\mathcal{P}_\ell$ | $f$ | Probability that SU decodes the PU's feedback message |
| $D_\text{p,nc}$ | Average queueing delay at the PU queue with no cooperation | $\mu_\text{p,nc}$ | Average service rate of the PU queue with no cooperation |
| $\mathcal{E}_\ell$ | Secondary mean transmit energy under scheme $\mathcal{P}_\ell$ | $E$ | Maximum transmit energy by the SU |
| $b$ | PU packet size in bits | $T_i$ and $W_i$ | Time and bandwidth assigned to user $i \in \{\text{p}, \text{s}\}$ under cooperation |

Both SU and PU transmit with a fixed power spectral density of $P_\text{o}$ Watts/Hz. The total transmit power changes based on the used bandwidth per transmission. When a node transmit over a bandwidth of $W_j$ Hz, the average transmit power is $P_\text{o} W_j$ Watts. Time is slotted and a slot has a duration of $T$ seconds. Channel coefficient between node j and node k, denoted by $\zeta_\text{j,k}$, is distributed according to a circularly symmetric Gaussian random variable, which is constant over one slot, but changes independently from one time slot to another. The expected value of the channel gain $\alpha_\text{j,k} = |\zeta_\text{j,k}|^2$ is $\sigma_\text{j,k}^2$, where $|\cdot|$ denotes the magnitude of a complex argument. Each receiving signal is perturbed by a zero-mean additive white Gaussian noise (AWGN) with power spectral density $\mathcal{N}_\text{o}$ Watts/Hz. The outage of a channel (link) occurs when the transmission rate exceeds the channel rate. The outage probability between two nodes j and k without and with the presence of interference from other nodes are denoted by $\mathbb{P}_\text{j,k}$ and $\mathbb{P}_\text{j,k}^{(\mathcal{I})}$, respectively. These outage probabilities are functions of the number of bits in a data packet, the slot duration, the transmission bandwidth, the transmit powers, and the average channel gains as detailed in Appendices A and B.

*B. Primary Access and Secondary Access Permission*

The PU transmits its data whenever it has a packet to send. That is, it does not have any restrictions on using the spectrum. Without cooperation, the PU uses the entire time slot duration and total bandwidth for its own data signal transmissions, while the SU does not gain any spectrum/channel access even if the PU's queue is empty. This is because, in practice, the SU may erroneously misdetect the primary activity and hence it may cause harmful interruption on the primary system operation, e.g., collisions and packets loss, that can cause sever packet losses and data delays. In case of cooperation, and based on the proposed cooperative cognitive schemes that will be explained shortly, the PU will release a portion of its time slot duration and total bandwidth to the SU. The SU will then be allowed to use the spectrum. In practice, the SU may get permission to access the spectrum if it either provides economic incentives for the PU or performance enhancement incentives. Similar to [5], [14], [17] and the references therein, we consider performance enhancement incentives.

### III. QUEUE STABILITY, PU QUEUE MODEL, AND PU QUEUEING DELAY

*A. Stability*

A queueing system is said to be stable if its size is bounded all the time. More specifically, let $Q^\mathbb{T}$ denote the length of

---
[3]This will be discussed later in Appendix B.

queue $Q$ at the beginning of time slot $\mathbb{T} \in \{1,2,3,\dots\}$. Queue $Q$ is said to be stable if

$$\lim_{x\to\infty}\lim_{\mathbb{T}\to\infty}\Pr\{Q^{\mathbb{T}}<x\}=1 \qquad(1)$$

For the PU queue, we adopt a late-arrival model where a newly arrived packet to the queue is not served in the arriving time slot even if the queue is empty.[4] Let $\mathcal{A}_{\rm p}^{\mathbb{T}}$ denote the number of arrivals to queue $Q_{\rm p}$ in time slot $\mathbb{T}$, and $\mathcal{H}_{\rm p}^{\mathbb{T}}$ denote the number of departures from queue $Q_{\rm p}$ in time slot $\mathbb{T}$. The queue length evolves according to the following form:

$$Q_{\rm p}^{\mathbb{T}+1}=(Q_{\rm p}^{\mathbb{T}}-\mathcal{H}_{\rm p}^{\mathbb{T}})^{+}+\mathcal{A}_{\rm p}^{\mathbb{T}} \qquad(2)$$

where $(z)^{+}$ denotes $\max(z,0)$. We assume that departures occur before arrivals, and the queue size is measured at the early beginning of the time slot [21].

*B. PU Queueing Delay*

Let $\mu_{\rm p}=\tilde{\mathcal{H}}_{\rm p}$, where $\tilde{\mathcal{V}}$ denotes the expected value of $\mathcal{V}$, be a general notation for the mean service rate of the PU queue. Solving the state balance equations of the Markov chain modeling the PU queue (Fig. 1), it is straightforward to show that the probability that the PU queue has $m\geq 1$ packets, denoted by $0\leq \nu_m \leq 1$, is given by

$$\nu_m=\frac{\nu_0}{\overline{\mu_{\rm p}}}\left(\frac{\lambda_{\rm p}\overline{\mu_{\rm p}}}{\overline{\lambda_{\rm p}}\mu_{\rm p}}\right)^m=\frac{\nu_0}{\overline{\mu_{\rm p}}}\eta^m,\ m=1,2,\dots,\infty \qquad(3)$$

where $\eta=\frac{\lambda_{\rm p}\overline{\mu_{\rm p}}}{\overline{\lambda_{\rm p}}\mu_{\rm p}}$. Since the sum over all states' probabilities is equal to one, i.e., $\sum_{m=0}^{\infty}\nu_m=1$, the probability of the PU queue being empty is obtained by solving the following equation

$$\nu_0+\sum_{m=1}^{\infty}\nu_m=\nu_0+\nu_0\sum_{m=1}^{\infty}\frac{1}{\overline{\mu_{\rm p}}}\eta^m=1 \qquad(4)$$

After some mathematical manipulations and simplifications, $\nu_0$ is given by

$$\nu_0=1-\frac{\lambda_{\rm p}}{\mu_{\rm p}} \qquad(5)$$

The PU queue is stable if $\mu_{\rm p}>\lambda_{\rm p}$. Applying Little's law, the PU average queueing delay, denoted by $D_{\rm p}$, is then given by

$$D_{\rm p}=\frac{1}{\lambda_{\rm p}}\sum_{m=0}^{\infty}m\nu_m \qquad(6)$$

Using (3), $D_{\rm p}$ is rewritten as

$$D_{\rm p}=\frac{\nu_0}{\lambda_{\rm p}\overline{\mu_{\rm p}}}\sum_{m=1}^{\infty}m\eta^m \qquad(7)$$

Substituting with $\nu_0$ into $D_{\rm p}$, the PU average queueing delay is then given by

$$D_{\rm p}=\frac{1-\lambda_{\rm p}}{\mu_{\rm p}-\lambda_{\rm p}} \qquad(8)$$

Following are some important remarks. Firstly, the PU average queueing delay cannot be less than one time slot, which is attained when the denominator of (8) equals to the numerator. This condition implies that $\mu_{\rm p}=1$ packets/time slot, i.e., the minimum of $D_{\rm p}$ is attained if the service rate of the PU queue is equal to unity.

---
[4]This queueing model is considered in many papers, see for example, [10], [15], [21] and the references therein.

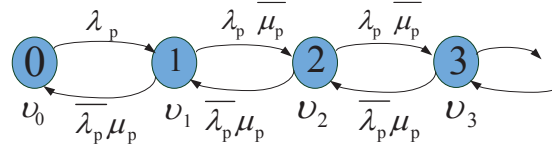

Fig. 1: Markov chain of the PU's queue. State self-transitions are omitted for visual clarity.

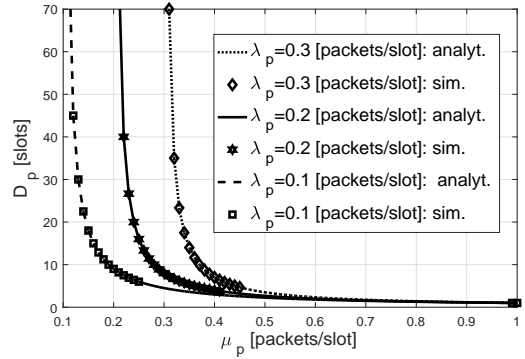

Fig. 2: PU average queueing delay versus $\mu_{\rm p}$ for different values of $\lambda_{\rm p}$.

To verify the average queueing delay expression and show the impact of both $\lambda_{\rm p}$ and $\mu_{\rm p}$, we plotted the curves in Fig. 2. As shown in the figure, increasing $\mu_{\rm p}$ decreases the average queueing delays. Moreover, the average queueing delay is increasing with the increase of the data arrival rate $\lambda_{\rm p}$. As shown analytically, the minimum average queueing delay is 1 time slot when $\mu_{\rm p}=1$ packets/slot.

Secondly, the primary packets average queueing delay, $D_{\rm p}$, decreases with increasing of the mean service rate of the PU queue, $\mu_{\rm p}$. On the other hand, $\mu_{\rm p}$ depends on the channels outage probabilities which, in turn, are functions of the links' parameters, packet size, transmission time durations, occupied bandwidth, and many other parameters as shown in Appendices A and B.

Hereinafter, when necessary, we append a second **subscript** to the used notations to distinguish between the cases of cooperation ('c') and no cooperation ('nc'). We also append a new **superscript** to distinguish between the proposed schemes.

## IV. NON-COOPERATIVE AND COOPERATIVE USERS

*A. Non-Cooperative Users*

Let $T$ denote the time slot duration that a PU is allowed to transmit data over a total bandwidth of $W$ Hz. Without cooperation, the time slot is divided into two non-overlapped phases: a transmission data phase, which takes place over the time interval $[0, T-\tau_f]$; and a feedback phase whose length is $\tau_f$ seconds, which takes place over the time interval $[T-\tau_f, T]$. The feedback phase is used by the primary destination to notify the primary transmitter about the decodability status of its packet. If the PU queue is nonempty, the PU transmits exactly one packet of size $b$ bits to its respective destination. The PU and primary destination implement an Automatic Repeat-reQuest (ARQ) error control scheme. The primary destination uses the cyclic redundancy code (CRC) bits attached to each packet to ascertain the decodability status of the received packet. The retransmission process is based on an acknowledgment/negative-acknowledgement (ACK/NACK) mechanism, in which short-length packets are broadcasted by the primary destination to inform the primary transmitter about its packet reception status. If the PU receives an ACK over the time interval $[T - \tau_f, T]$, it removes the data packet stored at the head of its queue; otherwise, a retransmission of the packet is generated at the following time slot(s). The ARQ scheme is untruncated which means that there is no maximum on the number of retransmissions and an erroneously received packet is retransmitted until it is decoded correctly at the primary destination [10], [13], [15], [22].

Without cooperation, a data packet at the head of the PU queue is served if the p → pd link is not in outage. Using the derived results in Appendix A for the channel outage probability, the mean service rate of the PU queue, denoted by $\mu_{\rm p,nc}$, is given by

$$\mu_{\rm p,nc} = \exp\left(-\mathcal{N}_\circ \frac{2^{\frac{b}{W(T-\tau_f)}}-1}{P_\circ \sigma_{\rm p,pd}^2}\right) \qquad (9)$$

It is noteworthy from (9) that increasing the feedback duration, $\tau_f$, decreases the service rate of the PU queue. This is because the time available for transmission data decreases with increasing $\tau_f$; hence, the outage probability increases which reduces the service rate. Since the PU transmits with a fixed rate of $\mathcal{R}_{\rm p} = \frac{b}{W(T-\tau_f)}$ bits per channel use, increasing $W$ or $T$ decreases the channel outage probability as seen in (9). However, increasing $\mathcal{R}_{\rm p}$ decreases the throughput since the number of decoded bits per seconds is decreased. Hence, one should compute the number of decoded bits per second per Hz which is given by

$$\mu_{\rm p,nc} = \exp\left(-\mathcal{N}_\circ \frac{2^{\frac{b}{WT(1-\frac{\tau_f}{T})}}-1}{P_\circ \sigma_{\rm p,pd}^2}\right)\frac{b}{WT} \qquad (10)$$

Letting $\mathcal{R}_{\rm p} = \frac{b}{WT}$, we have

$$\mu_{\rm p,nc} = \exp\left(-\mathcal{N}_\circ \frac{2^{\frac{\mathcal{R}_{\rm p}}{(1-\frac{\tau_f}{T})}}-1}{P_\circ \sigma_{\rm p,pd}^2}\right)\mathcal{R}_{\rm p} \qquad (11)$$

Using the first derivative of $\mu_{\rm p,nc}$ in (10) with respect to $b$, the optimal packet size is

$$b^\star = WT(1 - \frac{\tau_f}{T})\frac{\mathcal{W}\left(\frac{P_\circ \sigma_{\rm p,pd}^2}{\mathcal{N}_\circ}\right)}{\ln(2)} \qquad (12)$$

where $\mathcal{W}(\cdot)$ is Lambert-$\mathcal{W}$ (omega) function. From this interesting result, increasing the feedback duration $\tau_f$ will decrease the packet size. This is expected since the allowed time to send a data packet will decrease. On the other hand, we can see that increasing the time slot duration $T$ or the average receive SNR $\frac{P_\circ \sigma_{\rm p,pd}^2}{\mathcal{N}_\circ}$ at the primary destination increases the optimal packet size. This implies that more packet size can be supported by the communication system. However, increasing $T$ and $W$ linearly increase the optimal packet size. When $T$

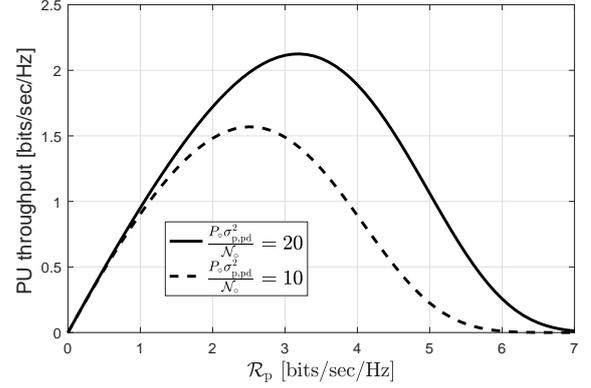

Fig. 3: PU throughput [bits/sec/Hz] versus $\mathcal{R}_{\rm p}$ [bits/sec/Hz].

is sufficiently longer than $\tau_f$, this leads to

$$b^\star = WT \frac{\mathcal{W}\left(\frac{P_\circ \sigma_{\rm p,pd}^2}{\mathcal{N}_\circ}\right)}{\ln(2)} \qquad (13)$$

Thus, the number of bits per channel use $\mathcal{R}_{\rm p}$ that maximizes the throughput (in bits/sec/Hz) is

$$\mathcal{R}_{\rm p}^\star = \frac{b^\star}{WT} = \frac{\mathcal{W}\left(\frac{P_\circ \sigma_{\rm p,pd}^2}{\mathcal{N}_\circ}\right)}{\ln(2)} \qquad (14)$$

To verify our analytical finding and show the impact of $\mathcal{R}_{\rm p}$ on the PU throughput [bits/sec/Hz], we plot Fig. 3. As can be seen from Fig. 3, the PU throughput increases with $\mathcal{R}_{\rm p}$ until a peak is reached, then the throughput decreases until it reaches zero. Hence, there is an optimal value for the packet size ($b^\star$ or $\mathcal{R}_{\rm p}^\star$ for a given $TW$) that maximizes the PU throughput. This value is given by (14). Increasing the average receive SNR $P_\circ \sigma_{\rm p,pd}^2 \mathcal{N}_\circ$ increases the PU throughput and also increases the optimal $\mathcal{R}_{\rm p}^\star$. This matches our discussion below (12).

According to (8), and using (9), the PU average queueing delay in case of non-cooperative PU is given by

$$D_{\rm p,nc} = \frac{1-\lambda_{\rm p}}{\mu_{\rm p,nc}-\lambda_{\rm p}} = \frac{1-\lambda_{\rm p}}{\exp\left(-\mathcal{N}_\circ \frac{2^{\frac{b}{W(T-\tau_f)}}-1}{P_\circ \sigma_{\rm p,pd}^2}\right) - \lambda_{\rm p}} \qquad (15)$$

with $\lambda_{\rm p} < \mu_{\rm p,nc}$ which represents the stability condition of the PU queue when there is no cooperation.

*B. Cooperative Users*

When the SU is able to assist the PU with relaying a portion of the primary packets, the PU, in return, may release a portion of its spectrum to the SU for its own transmission data if cooperation is beneficial for the PU. In addition to releasing some bandwidth for the SU, the PU releases a portion of its time slot duration to the SU to retransmit the primary packet. If the cooperation is beneficial for the PU, it cooperates with the SU. If the PU queue is nonempty, the PU releases $W_{\rm s} \leq W$ Hz to the SU for its own data transmission, and releases $T_{\rm s}$ seconds of the time slot to the SU for relaying the primary packets. The used bandwidth for both transmission and retransmission of the primary packet is $W_{\rm p} = W - W_{\rm s}$ Hz with transmission times $T_\text{p}$ and $T_\text{s}$, respectively. Throughout the paper, we use the analogy of subbands to distinguish between the primary operational frequency subband, $W_\text{p}$, and the secondary operational frequency subband, $W_\text{s}$.

*1) Spectrum Sensing:* The SU senses the primary subband, $W_\text{p}$, for $\tau_\text{s}$ seconds from the beginning of the time slot to detect the possible activities of the PU. If this subband is sensed to be idle (unutilized by the PU), the SU exploits its availability by sending some of its data bits. We assume that the SU employs an energy-detection spectrum-sensing algorithm. Specifically, the SU collects a number of samples over a time duration $\tau_\text{s} \ll T$, measures their energy, and then compares the measured energy to a predefined threshold to make a decision on the PU activity [23]. Detection reliability and quality depend on the sensing duration, $\tau_\text{s}$, and can be enhanced by increasing $\tau_\text{s}$. Specifically, as $\tau_\text{s}$ increases, the primary detection becomes more reliable at the expense of reducing the time available for secondary transmission over the primary subband if the PU is actually inactive. This is the essence of the sensing-throughput tradeoff in cognitive radio systems [23].

Since the sensing outcome is imperfect and subject to errors due to AWGN, the SU may interfere with the PU and cause some packet loss and collisions. To capture the impact of sensing errors, we define $P_\text{MD}$ as the probability of misdetecting the primary activity by the secondary terminal, which represents the probability of considering the PU inactive while it is actually active; and $P_\text{FA}$ as the probability that the sensor of the secondary terminal generates a false alarm, which represents the probability of considering the PU active while it is actually inactive. The values of sensing errors probabilities are derived in Appendix C.

*2) Important Notes and Remarks:* In the following, we state some important notes regarding our proposed cooperative schemes.

- A communication link is assumed to be 'ON' in a given time slot if it is not in outage. In particular, a link is ON if the instantaneous data rate of that link is higher than the used transmission data rate at the transmitter. In this case, the probability of bit-error rate is very low and can be neglected. Otherwise, the communication link is said to be 'OFF' (i.e., unable to support the transmission rate). In other words, the bit-error rate is unbounded (average symbol error rate is almost 1) and data retransmission should take place in the following transmission times.
- The CSI of the $\text{s} \to \text{pd}$, $\text{p} \to \text{s}$ and $\text{s} \to \text{sd}$ links are assumed to be known accurately at the SU (a similar assumption of knowing the CSI at the transmitters is found in many papers, for example, [14] and the references therein).[5] This allows the SU to better utilize the spectrum and helps the PU whenever necessary and possible.
- We assume that the SU always has data bits to transmit and it transmits its data with the instantaneous channel rate of its link, i.e., $\text{s} \to \text{sd}$ link. This is realized through the implementation of adaptive modulation schemes which is one of the main advantage of the cognitive radio devices [14].
- Since the SU has the CSI of all the communication links as explained in the previous bullet, in each time slot, the SU ascertains the state of the $\text{s} \to \text{pd}$ link, i.e., ON or OFF link, by comparing $\alpha_\text{s,pd}$ to the decoding threshold $\alpha_\text{th,s,pd}$. Further details on a link state is provided in Appendix A. After that, the SU can take decisions based on the other links to better help the PU.
- Since the SU operation is based on the spectrum sensing outcomes, the time assigned to channel sensing, denoted by $\tau_\text{s}$, should be less than the PU transmission time $T_\text{p}$ (i.e., $\tau_\text{s} < T_\text{p}$). In particular, the SU cannot set $\tau_\text{s}$ to be longer than the time assigned to PU transmission.
- If the $\text{p} \to \text{s}$ link is in outage (i.e., OFF), this means that the SU will not be able to decode the PU packet since the noise signal dominates the data signal and the transmission data rate is higher than the channel rate.
- Each PU packet comes with a CRC so that receivers (primary destination and SU) check the checksum to indicate the status of the decoded packet. Hence, if the SU cannot decode the primary packet in a time slot, i.e., the $\text{p} \to \text{s}$ link is in outage, or if the PU's queue is empty, the SU will not waste energy in forwarding what it receives from the wireless channel because it knows with certainty that the received packet is a noisy packet (i.e., has no data when the PU queue is empty). Consequently, the SU saves its energy from being wasted in a useless primary data retransmission, and it instead exploits that amount of energy for the transmission of its own data. This is critical since the SU energy is constrained and needs to be optimized.
- The data signals transmitted over subband $W_\text{s}$ are independent of the data signals transmitted over subband $W_\text{p}$. Hence, when there is an interference over subband $W_\text{p}$ due to simultaneous transmissions from the SU and the PU, the data signals over subband $W_\text{s}$ do not get affected.
- If the PU is active in a given time slot and the SU misdetects its activity, a concurrent transmission takes place over the primary subband, $W_\text{p}$. Hence, the SU data bits transmitted over $W_\text{p}$ are lost since the transmission data rate is higher than the link rate, and the primary packet could survive if the received SINR is higher than the decoding threshold. This event occurs with probability $\mathbb{P}_\text{p,pd}^{(\mathcal{I})}$.[6] See Appendix B for further details.
- We assume that the primary ARQ feedback is unencrypted and is available to the SU. A similar assumption is found in many references, e.g., [10] and the references therein.
- If the SU transmits concurrently with the primary destination during the feedback phase, the feedback message (packet) may be undecodable at the PU. For this reason, the SU remains silent/idle during the primary feedback duration to avoid disturbing the primary system operation.

---

[5]Note that the channel coefficient between the SU and the primary destination can be estimated by the primary destination and fed back to the SU. The primary destination only needs to send the state of the channel, i.e., ON or OFF, which can be realized through a one-bit binary feedback pilot signal.

[6]Throughout this paper, $\overline{\mathcal{X}} = 1 - \mathcal{X}$.

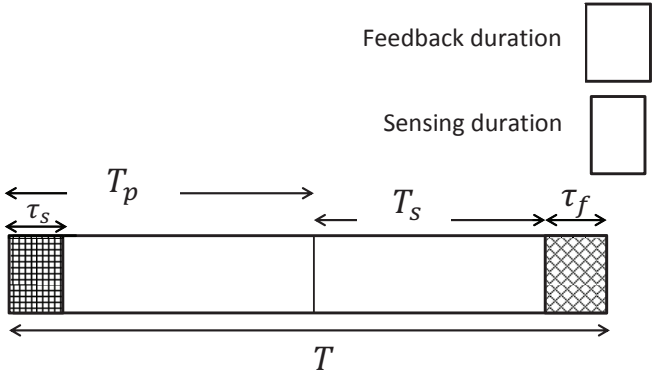

Fig. 4: Time slot structure under proposed scheme $\mathcal{P}_1$. In the figure, $\tau_s$ is the spectrum sensing time duration, $T_p$ is the PU transmission time of the primary data packet, $T_s$ is the time duration assigned to the secondary transmission of the primary packet, and $\tau_f$ is the feedback duration. Note that $T_p + T_s + \tau_f = T$.

## V. First Proposed Scheme

In this section, we explain our first proposed cooperative scheme, denoted by $\mathcal{P}_1$, and derive the achievable data rates and the energy emitted by the SU. The time slot structure under $\mathcal{P}_1$ is shown in Fig. 4. In our first proposed cooperative scheme, the operation of the SU during any arbitrary time slot changes over **four** phases: $[0, \tau_s]$, $[\tau_s, T_p]$, $[T_p, T_p + T_s]$, and $[T_p + T_s, T_p + T_s + \tau_f]$ (or simply $[T - \tau_f, T]$).

### A. Scheme Description

Before proceeding to the scheme description, we note that if the PU is active during a time slot, its transmission takes place over $[0, T_p]$, whereas the secondary retransmission of the primary packet takes place over $[T_p, T_p + T_s]$. The operation of the SU during each phase is described as follows.

*1) Time interval $[0, \tau_s]$:* The SU simultaneously senses the primary subband, $W_p$, and transmits its own data over $W_s$. The sensing outcome is then used for the secondary operation over $[\tau_s, T_p]$.

*2) Time interval $[\tau_s, T_p]$:* If the SU detects the PU to be active, it simultaneously transmits its own data over $W_s$, and attempts to decode the PU transmission over $W_p$. If the SU detects the PU to be inactive, it transmits its own data over both subbands, $W_p$ and $W_s$. If the PU is active and the SU finds the primary subband to be free of the PU transmission, there will be interference between the PU and the SU over $W_p$.

*3) Time interval $[T_p, T_p + T_s]$:* If the PU's queue is empty, the SU transmits its own data over both subbands. If the links $\text{p} \to \text{s}$ and $\text{s} \to \text{pd}$ are simultaneously ON and the PU queue is nonempty, the SU simultaneously transmits its own data over $W_s$ and retransmits the primary packet over $W_p$. If either the $\text{p} \to \text{s}$ link or the $\text{s} \to \text{pd}$ link is OFF, the SU transmits its own data over both subbands.

*4) Time interval $[T - \tau_f, T]$:* If the PU was active during $[0, T_p]$, then its respective receiver broadcasts a feedback message to indicate the status of the packet decodability.

Hence, the SU transmits its own data over $W_s$ and remains silent over $W_p$ to avoid causing any interference or disturbance for the feedback message transmission. If the PU was inactive during $[0, T_p]$, there is no feedback message in the current time slot. However, since the SU does not know the exact state of the PU during a time slot, it remains idle.

To summarize, the SU does not access the spectrum allocated to the PU, $W_p$, during the feedback duration to avoid disturbing the feedback message transmission.

### B. PU and SU Data Rates and SU Emitted Energy

A packet at the head of the PU queue $Q_{\text{p,c}}^{(1)}$ is served if the SU detects the primary activity correctly and either the direct path or the relaying path[7] is not in outage; or if the SU misdetects the primary activity and the $\text{p} \to \text{pd}$ link is not in outage. Let $\mu_{\text{p,c}}^{(\ell)}$ denote the mean service rate of the PU under scheme $\mathcal{P}_\ell$, $\ell \in \{1, 2\}$. The mean service rate of the PU queue under scheme $\mathcal{P}_1$ is then given by

$$\mu_{\text{p,c}}^{(1)} = \overline{P_{\text{MD}}}\bigg(1 - \mathbb{P}_{\text{p,pd}}\Big(1 - \overline{\mathbb{P}_{\text{p,s}}}\,\overline{\mathbb{P}_{\text{s,pd}}}\Big)\bigg) + P_{\text{MD}}\bigg(1 - \mathbb{P}_{\text{p,pd}}^{(\mathcal{I})}\bigg) \quad (16)$$

where $P_{\text{MD}}(1 - \mathbb{P}_{\text{p,pd}}^{(\mathcal{I})})$ denotes the probability of correct primary packet decoding at the primary destination when the SU misdetects the primary activity over $W_p$.

Let $\mathcal{R}_{\text{e}}^{(\ell)}$ and $\mathcal{R}_{\text{b}}^{(\ell)}$ denote the SU transmission data rate under scheme $\mathcal{P}_\ell$ when the PU queue is empty and nonempty, respectively, and $R = \log_2\left(1 + \frac{\alpha_{\text{s,sd}} P_\circ}{\mathcal{N}_\circ}\right)$ denote the instantaneous data rate of the $\text{s} \to \text{sd}$ link in bits/sec/Hz.

Based on the description of scheme $\mathcal{P}_1$, the SU transmission data rate when the PU queue is empty is given by

$$\mathcal{R}_{\text{e}}^{(1)} = \bigg(\tau_s \delta_s + (T_p - \tau_s)(P_{\text{FA}} \delta_s + \overline{P_{\text{FA}}}) + T_s\bigg) W R \quad (17)$$

where $\delta_s = W_s/W$. When the PU queue is nonempty, the SU transmission data rate is given by

$$\mathcal{R}_{\text{b}}^{(1)} = \bigg(\Big(T_p \delta_s + (P_{\text{MD}} + \overline{P_{\text{MD}}} \mathbb{P}_{\text{p,s}}) T_s \\ + \overline{P_{\text{MD}}} \overline{\mathbb{P}_{\text{p,s}}} T_s (\overline{\mathbb{P}_{\text{s,pd}}} \delta_s + \mathbb{P}_{\text{s,pd}})\Big)\bigg) W R \quad (18)$$

The term $\mathbb{P}_{\text{p,s}}$ appears in $\mathcal{R}_{\text{b}}^{(1)}$ because the SU, when the $\text{p} \to \text{s}$ link is in outage, uses the entire bandwidth for its own data transmission. Furthermore, the term $\mathbb{P}_{\text{s,pd}}$ appears in the expression of $\mathcal{R}_{\text{b}}^{(1)}$ because the SU, in each time slot, knows the channel state between itself and the primary destination and uses the allocated bandwidth to the PU for its own transmission data when that channel is in outage.

Let $\mathbb{I}[L]$ denote the indicator function, where $\mathbb{I}[L] = 1$ if the argument is true. The SU transmission data rate when it operates under scheme $\mathcal{P}_\ell$ is then given by

$$\mathcal{R}_{\text{s}}^{(\ell)} = \mathbb{I}[Q_{\text{p,c}}^{(\ell)} = 0] \mathcal{R}_{\text{e}}^{(\ell)} + \mathbb{I}[Q_{\text{p,c}}^{(\ell)} \neq 0] \mathcal{R}_{\text{b}}^{(\ell)} \quad (19)$$

---

[7]The relaying path is defined as the path connecting the PU to primary destination through the SU; namely, links $\text{p} \to \text{s}$ and $\text{s} \to \text{pd}$. Since the channels are independent, the probability of the relaying path being not in outage is $\overline{\mathbb{P}_{\text{p,s}}}\,\overline{\mathbb{P}_{\text{s,pd}}}$.

The expected value of $\tilde{\mathbb{I}}[L]$ is equal to the probability of the argument event. That is,

$$\tilde{\mathbb{I}}[L] = \Pr\{L\} \tag{20}$$

The mean SU transmission data rate is then given by

$$\tilde{\mathcal{R}}_{\rm s}^{(\ell)} = \Pr\{Q_{\rm p,c}^{(\ell)} = 0\}\tilde{\mathcal{R}}_{\rm e}^{(\ell)} + \Pr\{Q_{\rm p,c}^{(\ell)} \neq 0\}\tilde{\mathcal{R}}_{\rm b}^{(\ell)} \tag{21}$$

Recalling that $\Pr\{Q_{\rm p,c}^{(\ell)} = 0\} = \nu_{0,\rm c}^{(\ell)}$ and $\Pr\{Q_{\rm p,c}^{(\ell)} \neq 0\} = 1 - \nu_{0,\rm c}^{(\ell)}$, the mean SU transmission data rate under scheme $\mathcal{P}_1$ is then given by

$$\begin{aligned}\tilde{\mathcal{R}}_{\rm s}^{(1)} = &\nu_{0,\rm c}^{(1)}\Big(\tau_{\rm s}\delta_{\rm s} + (T_{\rm p}-\tau_{\rm s})(P_{\rm FA}\delta_{\rm s} + \overline{P_{\rm FA}}) + T_{\rm s}\Big)W\mathcal{G}_{\rm s} \\ &+ \overline{\nu_{0,\rm c}^{(1)}}\Big(T_{\rm p}\delta_{\rm s} + (P_{\rm MD} + \overline{P_{\rm MD}}\mathbb{P}_{\rm p,s})T_{\rm s} \\ & \qquad\qquad + \overline{P_{\rm MD}}\mathbb{P}_{\rm p,s}T_{\rm s}(\overline{\mathbb{P}_{\rm s,pd}}\delta_{\rm s} + \mathbb{P}_{\rm s,pd})\Big)W\mathcal{G}_{\rm s}\end{aligned} \tag{22}$$

where $\mathcal{G}_{\rm s}$ is the expected value of $\log_2(1 + \alpha_{\rm s,sd}\frac{P_{\circ}}{\mathcal{N}_{\circ}})$, which is given by (see Appendix D for details)

$$\mathcal{G}_{\rm s} = \frac{1}{\ln(2)}\exp\left(\frac{1}{\frac{P_{\circ}}{\mathcal{N}_{\circ}}\sigma_{\rm s,sd}^2}\right)\Gamma\left(0, \frac{1}{\frac{P_{\circ}}{\mathcal{N}_{\circ}}\sigma_{\rm s,sd}^2}\right) \tag{23}$$

where $\Gamma(m,s) = \int_{1/s}^{\infty}\exp(-z)z^{m-1}dz$ is the upper incomplete Gamma function.

According to the described scheme, the mean SU transmit energy, denoted by $\mathcal{E}_1$, is given by

$$\begin{aligned}\mathcal{E}_1 = &\nu_{0,\rm c}^{(1)}\Big(\tau_{\rm s}\delta_{\rm s} + (T_{\rm p}-\tau_{\rm s})(P_{\rm FA}\delta_{\rm s} + \overline{P_{\rm FA}}) + T_{\rm s}\Big)WP_{\circ} \\ &+ \overline{\nu_{0,\rm c}^{(1)}}\Big(\tau_{\rm s}\delta_{\rm s} + (T_{\rm p}-\tau_{\rm s})(\overline{P_{\rm MD}}\delta_{\rm s} + P_{\rm MD}) + T_{\rm s}\Big)WP_{\circ}\end{aligned} \tag{24}$$

Note that we assume that the maximum average emitted secondary energy is $E$; hence, $\mathcal{E}_1$ must be at most $E$.

## VI. SECOND PROPOSED SCHEME

In our second scheme, denoted by $\mathcal{P}_2$, we assume a variation in the primary feedback mechanism to further improve the achievable performance for both PU and SU. More specifically, we assume the existence of two primary feedback phases within each time slot. Each transmission of the primary packet by either the PU or the SU is followed by a feedback phase to inform the transmitter (PU or SU) about the decodability of the transmitted packet. In other words, a feedback message is sent by the primary destination when it receives a copy of the expected primary packet.[8] The first feedback phase is preceded by the PU transmission of the primary packet, whereas the second feedback phase is preceded by the SU transmission of the primary packet. The PU queue drops the packet if it receives at least one ACK in any time slot. Otherwise, the packet will be retransmitted by the PU in the following time slots until its correct decoding at the primary destination.

On the one hand, the gain of this cooperative scheme over the first proposed scheme lies in its ability to prevent unnecessary retransmissions of a successfully decoded primary packet at the primary destination. More specifically, if the primary destination can decode the PU transmission correctly, then the SU does not need to retransmit the same primary packet over the primary subband and over the time assigned for relaying; hence, the SU can instead use the time assigned for relaying and the primary subband to transmit its own data bits to its destination.[9] Consequently, using scheme $\mathcal{P}_2$ enables the SU to increase its average transmission rate via using the allocated bandwidth and time duration for PU transmissions and its transmit energy to send its own data. On the other hand, there is a considerable cost due to appending an extra feedback duration to the time slot. This cost is converted to an increase in the outage probabilities of the links and the reduction in the users' rates. This is because the total time allocated for data bits and packets transmissions is reduced by $\tau_f$ seconds relative to the total transmission time in case of scheme $\mathcal{P}_1$.

Under cooperative scheme $\mathcal{P}_2$, the secondary operation in any arbitrary time slot changes over **five** phases as shown in Fig. 5: $[0, \tau_{\rm s}]$, $[\tau_{\rm s}, T_{\rm p}]$, $[T_{\rm p}, T_{\rm p}+\tau_f]$, $[T_{\rm p}+\tau_f, T_{\rm p}+\tau_f+T_{\rm s}]$ and $[T-\tau_f, T]$.

### A. Decoding of Primary Feedback Message at the SU

The correctness of the feedback message decoding at the SU is ascertained using the checksum appended to the feedback message packet. The decoding of a primary feedback message at the SU can be modeled as an erasure channel model. In particular, the primary feedback message is assumed to be decoded correctly at the SU with probability $f$. If the SU cannot decode the primary feedback message in a given time slot,[10] it considers this feedback message as a NACK feedback message. Another possibility is to assume that the SU considers the "nothing" as a NACK message with probability $\omega$ and considers it as an ACK message with probability $\overline{\omega}$. Using such parameter allows the SU to use a fraction of the "nothing" events that would be an ACK, which means that the SU does not need to retransmit the primary packet, for its own data bits transmission. The SU can optimize over $\omega$ to alleviate wasting the channel resources without further contribution to the primary service rate when the primary packet is already decoded successfully at the primary destination. The primary mean service rate in this case is given by

$$\mu_{\rm p,c}^{(2)} = \overline{P_{\rm MD}}\Big(1 - \mathbb{P}_{\rm p,pd}\Big(1 - \beta\overline{\mathbb{P}_{\rm p,s}}\,\overline{\mathbb{P}_{\rm s,pd}}\Big)\Big) + P_{\rm MD}\Big(1 - \mathbb{P}_{\rm p,pd}^{(\mathcal{I})}\Big) \tag{25}$$

where $\beta = f + \overline{f}\omega$ is the probability of considering the overheard feedback message as a NACK when the primary destination sends a NACK feedback (which occurs if the ${\rm p} \to {\rm pd}$ link is in outage). From (25), the primary mean service rate is parameterized by $\omega$. The maximum primary service rate is attained when $\omega = 1$ since the SU will relay more PU packets. For simplicity, we consider the case of $\omega=1$ which guarantees the highest QoS for the PU.

---

[8] Each packet comes with an identifier (ID) and a certain labeled number that is generated by the transmitter. In addition, the destination sends the expected number of the next packet as part of the feedback message.

[9] This is because the retransmission of the primary packet by the secondary transmitter does not provide further contribution to the primary throughput. In addition, the retransmission of the primary packet causes both energy and bandwidth losses that can be used otherwise for the SU data transmission.

[10] This event is referred to as "nothing" event. The "nothing" event is considered when the SU fails in decoding the feedback message, or when the PU is idle at this time slot, i.e., $Q_{\rm p} = 0$.

*B. Scheme Description*

The PU transmission occurs over $[0, T_p]$ and the secondary retransmission of a primary packet occurs over $[T_p+\tau_f, T_p+\tau_f+T_s]$. Note that the feedback message is considered by the SU as a NACK feedback message 1) if the p $\to$ pd link is in outage and the feedback message is decoded correctly at the SU terminal; or 2) if the feedback message is undecodable at the SU. The probability that the SU considers the overheard primary feedback message as a NACK is then given by

$$\Gamma_f = \mathbb{P}_{\mathrm{p,pd}} f + \overline{f} \tag{26}$$

In the sequel of this subsection, we describe the behavior of the SU during each phase.

*1) Time interval $[0, \tau_s]$ and $[\tau_s, T_p]$:* The operation of the system over the time intervals $[0, \tau_s]$ and $[\tau_s, T_p]$ is similar to the first cooperative scheme during the same time intervals.

*2) Time interval $[T_p, T_p+\tau_f]$:* If the PU queue is nonempty during the ongoing time slot, at the end of the PU dedicated transmission time, the SU transmits its own data over $W_s$, and remains silent over $W_p$ to avoid causing a concurrent transmission with the feedback message transmitted from the primary destination to the PU. If the PU queue is empty during the ongoing time slot, the SU transmits its own data over both subbands.

*3) Time interval $[T_p + \tau_f, T - \tau_f]$:* Upon decoding the entire primary packet, the SU discerns the actual (true) state of the PU, i.e., active or inactive. The SU transmits its own data over both subbands 1) if the PU was active during the time interval $[0, T_p]$, the primary destination correctly decoded the PU packet, and the SU successfully decoded the primary feedback message, i.e., considered it as an ACK feedback; or 2) if the s $\to$ pd link is in outage; or 3) if the PU was inactive during the time interval $[0, T_p]$. If the PU was active during the time interval $[0, T_p]$, the secondary terminal considered the feedback message sent over $[T_p, T_p+\tau_f]$ as a NACK feedback, and the s $\to$ pd link is not in outage; the SU simultaneously transmits its own data over $W_s$ and retransmits the primary packet over $W_p$.

*4) Time interval $[T - \tau_f, T]$:* If the SU retransmitted the packet over $[T_p + \tau_f, T - \tau_f]$, another feedback message will be sent over this phase by the primary destination. Hence, the SU simultaneously transmits its own data over $W_s$ and remains silent over $W_p$. If the SU decides not to retransmit the primary packet, there will be no primary feedback message. Therefore, the SU transmits its own data over both subbands. If the PU queue is empty during the ongoing time slot, the SU transmits its own data over both subbands over this feedback duration (i.e., $[T - \tau_f, T]$).

*C. PU and SU Data Rates and the SU Emitted Energy*

A data packet stored at the head of the PU queue $Q_{\mathrm{p,c}}^{(2)}$ is served in a given time slot 1) if the SU detects the primary activity correctly, and the direct link is not in outage; or 2) if the SU detects the primary activity correctly, and the direct link is in outage, the SU considers the primary feedback message as a NACK signal, and the relaying link is not in outage; or 3) if the SU misdetects the primary activity, and the direct link is not in outage. The mean service rate of the PU queue is similar to the first scheme and is given by

$$\mu_{\mathrm{p}}^{(2)} = \overline{P_{\mathrm{MD}}}\left(1 - \mathbb{P}_{\mathrm{p,pd}}\left(1 - \overline{\mathbb{P}_{\mathrm{p,s}}}\ \overline{\mathbb{P}_{\mathrm{s,pd}}}\right)\right) + P_{\mathrm{MD}}\left(1 - \mathbb{P}_{\mathrm{p,pd}}^{(\mathcal{I})}\right) \tag{27}$$

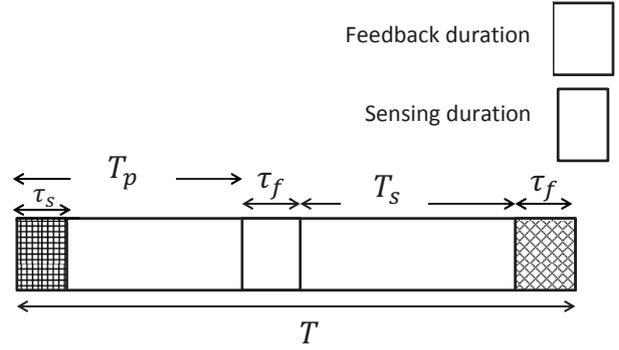

Fig. 5: Time slot structure under proposed scheme $\mathcal{P}_2$. In this scheme, there are two feedback message durations. Hence, $T_\mathrm{p} + T_\mathrm{s} + 2\tau_f = T$.

We note that the expression (27) is similar to (16). However, the maximum assigned transmission data times for users under $\mathcal{P}_2$ are lower than $\mathcal{P}_1$ as $\mathcal{P}_2$ has two feedback durations.

When the PU is inactive, the SU instantaneous transmission rate is given by

$$\mathcal{R}_{\mathrm{e}}^{(2)} = \left(\tau_\mathrm{s}\delta_\mathrm{s} + (T_\mathrm{p}-\tau_\mathrm{s})(\overline{P_{\mathrm{FA}}} + P_{\mathrm{FA}}\delta_\mathrm{s}) + T_\mathrm{s}\right)WR \tag{28}$$

When the PU is active, the SU instantaneous transmission rate is given by

$$\mathcal{R}_{\mathrm{b}}^{(2)} = \left(T_\mathrm{p}\delta_\mathrm{s} + \overline{P_{\mathrm{MD}}}T_\mathrm{s}\right.$$
$$\left. \times \left(\overline{\mathbb{P}_{\mathrm{p,s}}}\left(\Gamma_f\left(\overline{\mathbb{P}_{\mathrm{s,pd}}}\delta_\mathrm{s} + \mathbb{P}_{\mathrm{s,pd}}\right) + \overline{\Gamma}_f\right) + \mathbb{P}_{\mathrm{p,s}}\right) + P_{\mathrm{MD}}T_\mathrm{s}\right)WR \tag{29}$$

The mean SU transmission data rate is then given by

$$\tilde{\mathcal{R}}_{\mathrm{s}}^{(2)} = \nu_{0,\mathrm{c}}^{(2)}\left(\tau_\mathrm{s}\delta_\mathrm{s} + (T_\mathrm{p}-\tau_\mathrm{s})(\overline{P_{\mathrm{FA}}} + P_{\mathrm{FA}}\delta_\mathrm{s}) + T_\mathrm{s}\right)W\mathcal{G}_\mathrm{s}$$
$$+ \overline{\nu_{0,\mathrm{c}}^{(2)}}\left(T_\mathrm{p}\delta_\mathrm{s} + \overline{P_{\mathrm{MD}}}T_\mathrm{s}\right.$$
$$\left.\times \left(\overline{\mathbb{P}_{\mathrm{p,s}}}\left(\Gamma_f\left(\overline{\mathbb{P}_{\mathrm{s,pd}}}\delta_\mathrm{s} + \mathbb{P}_{\mathrm{s,pd}}\right) + \overline{\Gamma}_f\right) + \mathbb{P}_{\mathrm{p,s}}\right) + P_{\mathrm{MD}}T_\mathrm{s}\right)W\mathcal{G}_\mathrm{s} \tag{30}$$

According to the description of scheme $\mathcal{P}_2$, the mean SU transmit energy is given by

$$\mathcal{E}_2 = \nu_{0,\mathrm{c}}^{(2)}\left(\tau_\mathrm{s}\delta_\mathrm{s} + (T_\mathrm{p}-\tau_\mathrm{s})(\overline{P_{\mathrm{FA}}} + P_{\mathrm{FA}}\delta_\mathrm{s}) + T_\mathrm{s}\right)WP_\circ$$
$$+ \overline{\nu_{0,\mathrm{c}}^{(2)}}\left(\tau_\mathrm{s}\delta_\mathrm{s} + (T_\mathrm{p}-\tau_\mathrm{s})(\overline{P_{\mathrm{MD}}}\delta_\mathrm{s} + P_{\mathrm{MD}}) + T_\mathrm{s}\right)WP_\circ \tag{31}$$

## VII. Problem Formulation and Primary Mean Energy Savings

### A. Problem Formulation

We assume that users optimize over $T_p = T - \tau_f - T_s$ and $W_p = W - W_s$. It is noteworthy that there is a possibility to optimize over the spectrum sensing time $\tau_s$, however, for simplicity, we assume that the spectrum sensing time is fixed and predetermined. Sensing time optimization is out of scope of this paper. The optimization problem is formulated such that the secondary average data rate is maximized under a certain PU average queueing delay, the PU queue stability condition, and an energy constraint on the secondary average transmit energy, given by $\mathcal{E}_\ell \leq E$ (where $E$ denotes the maximum average SU transmit energy). The optimization problem under proposed scheme $\mathcal{P}_\ell \in \{\mathcal{P}_1, \mathcal{P}_2\}$ is stated as follows

$$\max_{T_p, W_p} \quad \tilde{\mathcal{R}}_s^{(\ell)}$$
$$\text{s.t.} \quad D_{p,c}^{(\ell)} < D_{p,nc}, \; \mu_{p,c}^{(\ell)} > \lambda_p, \; 0 \leq \mathcal{E}_\ell \leq E \quad (32)$$
$$\tau_s \leq T_p \leq \mathcal{T}^{(\ell)}, \; 0 \leq W_p \leq W, \; T_p + T_s = \mathcal{T}^{(\ell)}$$

where $\mathcal{T}^{(\ell)}$ is the operational constraint on $T_p + T_s$ when users operate under scheme $\mathcal{P}_\ell$, and $D_{p,c}^{(\ell)} = (1-\lambda_p)/(\mu_{p,c}^{(\ell)} - \lambda_p)$ is the average queueing delay of the PU queue under cooperation. Under our first cooperative scheme, the maximum allowable transmission time is $T - \tau_f$; hence, $\mathcal{T}^{(1)} = T - \tau_f$. On the other hand, under our second cooperative scheme, the maximum allowable transmission time is $T - 2\tau_f$; hence, $\mathcal{T}^{(2)} = T - 2\tau_f$. It should be pointed out here that if the primary feedback message is always undecodable at the SU, i.e., $f = 1$ or if the p $\to$ pd link is always in outage, scheme $\mathcal{P}_1$ always outperforms scheme $\mathcal{P}_2$. This is reasonable since the SU will always retransmit the primary packet with a lower transmission time for each user due to the existence of two feedback durations in $\mathcal{P}_2$. In addition, when $\tau_f$ increases, $\mathcal{P}_1$ may outperform $\mathcal{P}_2$ for some system parameters because it may be the case that the reduction in the maximum allowable transmission time due to the presence of an additional feedback duration is higher than the gain of knowing the status of the primary packet decodability at the SU before the secondary retransmission of the primary packet.

The optimization problem (32) is solved numerically using a two-dimensional grid-based search over $T_p$ and $W_p$. The optimal parameters obtained via solving the optimization problem (32) are announced to both users so that $W_p$ and $T_p$ are known at the PU and the SU before actual operation of the communications system. If the optimization problem is infeasible due to the dissatisfaction of one or more of the constraints in (32), the SU will not be allowed to use the spectrum and its achievable rate is zero. A simple method to solve the optimization problem (32) is to divide the domains of $T_p$ and $W_p$ into $K$ points. Then, solve the optimization problem (32) for $K^2$ times and select the solution that satisfies the constraints and has the highest objective function. Our proposed solution to the optimization problem in (32) is stated in Algorithm 1.

It is worth noting that the PU average queueing delay constraint can be replaced by a constraint on the mean service

---

**Algorithm 1** Optimization Procedure

1: Select a large number $K$
2: Set $i = 1$
3: *loop1*:
4: Generate $0 \leq \delta_p \leq 1$ where $\delta_p = W_p/W$
5: *loop2*:
6: Set $j = 1$
7: Generate $\frac{\tau_s}{T} \leq \Delta_p \leq \frac{\mathcal{T}^{(\ell)}}{T}$ where $\Delta_p = T_p/T$
8: Compute $W_s = W - T_p$ and $T_s = \mathcal{T}^{(\ell)} - T_p - \tau_s$
9: Compute $Z(i,j) = \tilde{\mathcal{R}}_s^{(\ell)}$ in (32)
10: Set $j = j + 1$
11: If $j \neq K$, **goto** *loop2*
12: Set $i = i + 1$
13: If $i \neq K$, **goto** *loop1*
14: Select $W_p = \delta_p W$ and $T_p = \Delta_p T$ that maximize $\tilde{\mathcal{R}}_s^{(\ell)}$ ($i$ and $j$ corresponding to highest $Z(i,j)$) and satisfy the constraints in (32)

---

rate of the PU queue. Since the delay constraint is given by $D_{p,c}^{(\ell)} = (1-\lambda_p)/(\mu_{p,c}^{(\ell)} - \lambda_p) < D_{p,nc} = (1-\lambda_p)/(\mu_{p,nc} - \lambda_p)$, the mean service rate of the PU queue under cooperation must be greater than the mean service rate of the PU queue without cooperation. In particular,

$$\mu_{p,c}^{(\ell)} > \mu_{p,nc} \quad (33)$$

Combining the delay constraint with the stability constraint, the PU queue mean service rate should be at least

$$\mu_{p,c}^{(\ell)} > \max\left\{\mu_{p,nc}, \lambda_p\right\} \quad (34)$$

### B. Mean Primary Energy Savings

In the absence of cooperation, the PU transmission takes place over $T - \tau_f$ seconds and occupies $W$ Hz. Hence, the PU energy consumption per time slot is $P_\circ W(T - \tau_f)$ joules/slot. However, when the SU helps the PU in relaying its packets, the PU transmits only in a fraction $T_p/T$ of the time slot with transmission bandwidth $W_p$ Hz. Hence, its energy consumption per time slot is only $P_\circ W_p T_p \leq P_\circ W(T - \tau_f)$ joules/slot. In this case, the average rate of the PU energy savings, defined as the ratio of the energy savings over the original energy consumption, is given by

$$\phi = \frac{P_\circ W(T - \tau_f)\Pr\{Q_{p,nc} \neq 0\} - P_\circ W_p T_p \Pr\{Q_{p,c}^{(\ell)} \neq 0\}}{P_\circ W(T - \tau_f)\Pr\{Q_{p,nc} \neq 0\}} \quad (35)$$

Using the fact that $\Pr\{Q_{p,nc} \neq 0\} = \lambda_p/\mu_{p,nc}$ if $\lambda_p < \mu_{p,nc}$, and 1 otherwise, $\Pr\{Q_{p,c}^{(\ell)} \neq 0\} = \lambda_p/\mu_{p,c}^{(\ell)}$, and noting that there is no cooperation if the PU queue is unstable, we get

$$\phi = 1 - \frac{W_p T_p}{W(T - \tau_f)} \frac{\max\{\mu_{p,nc}, \lambda_p\}}{\mu_{p,c}^{(\ell)}} \quad (36)$$

From the above ratio, we can see that the less the bandwidth and the transmission time that the PU occupies, the more energy savings for the PU. We note that the PU queue under cooperation should be stable, otherwise, the optimization problem is infeasible and there will be no cooperation. We also note that using less bandwidth and shorter transmission time improves the low probability of intercept/low probability of detection (LPD/LPI) characteristics of the communication link that appears to be especially critical in military applications. Hence, it is always useful to use shorter transmission times and lower bandwidth.

## VIII. NUMERICAL RESULTS

In this section, we present some simulations of the proposed cooperative schemes. We define a set of common parameters: the targeted false alarm probability is $P_{\text{FA}} = 0.1$, $W = 10$ MHz, $T = 5$ msec, $b = 5000$ bits, $E = 5 \times 10^{-6}$ joule, $\tau_s = 0.05T$, $\sigma^2_{s,pd} = \sigma^2_{s,sd} = 0.1$, $\sigma^2_{p,s} = 1$, $\mathcal{P}_o = 10^{-10}$ Watts/Hz, and $\mathcal{N}_o = 10^{-11}$ Watts/Hz. Fig. 6 shows the maximum average SU data rate of our proposed cooperative schemes. The second proposed scheme is plotted with three different values of $f$. The figure reveals the advantage of our second proposed scheme over our first proposed scheme for $f = 0.5$ and $f = 1$. However, for $f = 0$, the first proposed scheme outperforms the second one. This is reasonable since when $f = 0$ there is no gain from having a feedback message after the PU transmission; hence, using the second proposed scheme wastes $\tau_f$ seconds of the time slot that can be used otherwise in increasing users' data rates. The figure also demonstrates the impact of parameter $f$ on the performance of the second proposed scheme, i.e., scheme $\mathcal{P}_2$. As shown in the figure, increasing $f$ enhances the performance of scheme $\mathcal{P}_2$. In addition to the common parameters, the figure is generated using $\sigma^2_{p,pd} = 0.05$, $\tau_f = 0.05T$ and the values of $f$ in the figure's legend.

Fig. 7 shows the impact of the feedback message duration, $\tau_f$, on the performance of our proposed cooperative schemes. The mean SU transmission data rate and the PU data arrival rate feasible range decrease with increasing $\tau_f$. When the value of $\tau_f$ is considerable, i.e., $\tau_f = 0.2T$, the first scheme outperforms the second scheme. This is because the maximum allowable transmission data time of nodes under scheme $\mathcal{P}_2$ in this case is $T - 2\tau_f = 0.6T$, whereas the maximum allowable transmission time under scheme $\mathcal{P}_1$ is $T - \tau_f = 0.8T$. For small values of $\tau_f$, the second proposed scheme outperforms the first scheme since the SU can use the time duration assigned for relaying and the primary subband to transmit its data in case of correct packet decoding after the PU transmission. The parameters used to generate the figure are the common parameters, $\sigma^2_{p,pd} = 0.05$, $f = 1$ and the values of $\tau_f$ in the plot.

Figs. 8, 9 and 10 present the primary mean service rate, the PU average queueing delay, and the average PU power savings, respectively, under our proposed cooperative schemes. The case of non-cooperative users is also plotted in Figs. 8 and 9 for comparison purposes. The figures demonstrate the gains of the proposed schemes for the PU over the non-cooperation case. Note that without cooperation between the two users, the PU queue is unstable when $\lambda_{\text{p}} > 0.2$ packets/slot and, hence, the queueing delay is unbounded. On the other hand, with cooperation, the PU queue remains stable over the range from $\lambda_{\text{p}} = 0$ to $\lambda_{\text{p}} = 0.95$ packets/slot. The second scheme achieves better performance than the first scheme in terms of primary QoS. Fig. 10 reveals that more that 95% of the average primary energy will be saved for $\lambda_{\text{p}} = 0.2$ packets/slot. When $\lambda_{\text{p}} = 0.8$ packets/slot, the primary energy savings is almost 78%. For $\lambda_{\text{p}} \geq 0.95$, the PU queue becomes unstable even with cooperation; hence, the cooperation becomes non-beneficial for the PU and the PU ceases cooperation with the SU. Hence, the SU doest gain any access to the spectrum, and the primary energy savings becomes zero since the PU will send its data over the entire time slot duration and channel bandwidth. The parameters used to generate the figures are the common parameters, $\sigma^2_{p,pd} = 0.005$, $\sigma^2_{s,pd} = 1$, $\tau_f = 0.05T$ and $f = 1$. Note that the performance of our two proposed schemes are close to each other because the outage probability of the primary channel is high and the direct link (i.e., the p → pd link) is in outage most of the time. Hence, under scheme $\mathcal{P}_2$, the SU retransmits the primary packets almost every time slot instead of transmitting its own data signals. Accordingly, both proposed schemes almost achieve the same performance.

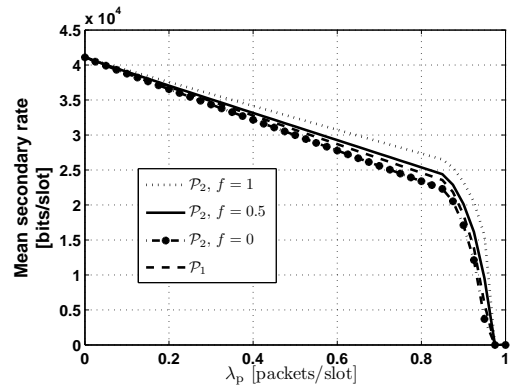

Fig. 6: The maximum SU data rate in bits per slot for the proposed schemes. Scheme $\mathcal{P}_2$ is plotted with three different values of primary feedback correct decoding, $f$.

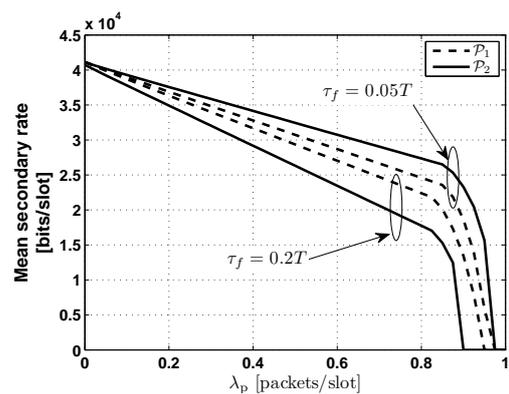

Fig. 7: The maximum mean SU data rate in bits per time slot. The schemes are plotted for two values of the feedback duration $\tau_f$.

## IX. CONCLUSIONS

In this paper, we developed two cooperative cognitive schemes which allow the SU to access the primary spectrum simultaneously with the PU. We showed the gains of our proposed cooperative schemes for the SUs and PUs. We also addressed the impact of the feedback process on users' data rates. Each of our proposed schemes can outperform the other for certain system parameters and they differ in terms of time slot structure. We showed that as the probability of feedback message decoding decreases, the second proposed scheme loses its advantage over the first proposed scheme. The PU energy savings under cooperation is more than $60\%$ for most of the PU packet arrival rate. Moreover, at low mean arrival rate at the PU data queue, the PU energy savings can be more than $95\%$. We also showed a significant reduction in the average queueing delay of the PU queue under cooperation relative to the no-cooperation case. As a future work, we can investigate the battery-based system where the communication nodes are equipped with rechargeable batteries with certain energy arrival rates.

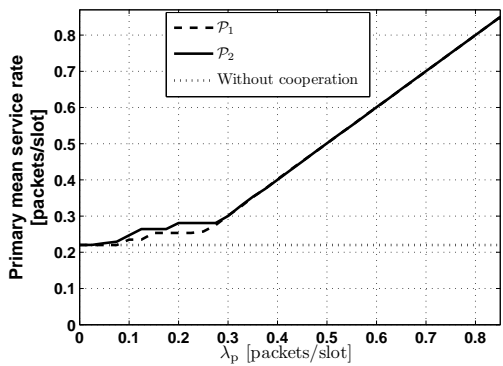

Fig. 8: The maximum mean primary stable throughput for the proposed schemes. The case of non-cooperative users is also plotted for comparison.

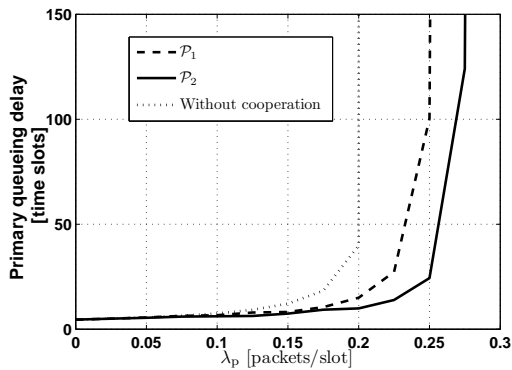

Fig. 9: The PU average queueing delay for the proposed schemes. The case of non-cooperative SU is also plotted for comparison purposes.

## APPENDIX A

In this Appendix, we present the outage probability expression of a link when the transmitter communicates with its respective receiver alone, i.e., without interference. Let $r_{j,k}$ be the transmission rate of node j while communicating with node k, $\gamma_{j,k}$ be the received SINR at node k when node j communicates with node k, and $\alpha_{j,k}$ be the associated channel gain with mean $\sigma_{j,k}^2$, which is exponentially distributed in the case of Rayleigh fading. The probability of channel outage between node j and node k is given by [24]

$$\mathbb{P}_{j,k} = \Pr\left\{r_{j,k} > \log_2\left(1 + \gamma_{j,k}\right)\right\} \quad (37)$$

where $\Pr\{\cdot\}$ denotes the probability of the event in the argument, and $\gamma_{j,k} = \frac{P_\circ \alpha_{j,k}}{\mathcal{N}_\circ}$. The formula (37) can be rewritten as

$$\mathbb{P}_{j,k} = \Pr\left\{\alpha_{j,k} < \frac{\mathcal{N}_\circ}{P_\circ}\left(2^{r_{j,k}} - 1\right)\right\} \quad (38)$$

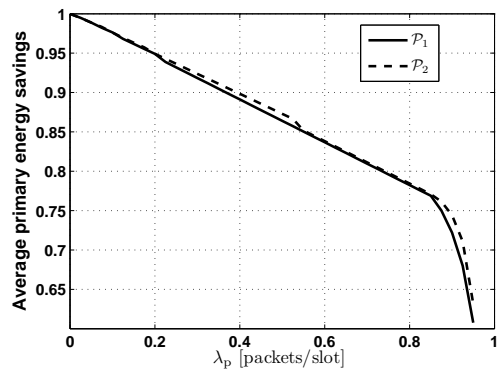

Fig. 10: PU power savings.

Let $\alpha_{\text{th},j,k} = \frac{\mathcal{N}_\circ}{P_\circ}(2^{r_{j,k}} - 1)$. We note that if $\alpha_{j,k} < \alpha_{\text{th},j,k}$, the channel is in outage (OFF), whereas if $\alpha_{j,k} \geq \alpha_{\text{th},j,k}$, the channel is not in outage (ON). It is worth pointing out here that increasing the transmission data time and the bandwidth assigned to any of the terminals decrease the outage probability, or equivalently increase the rate, of the link between that terminal and its respective receiver. That is, the outage probability of any of the links decreases exponentially with the increase of the transmission time and the bandwidth assigned to the transmitting node.

If the SU is available to assist, when the PU's queue is nonempty, the PU sends a packet of size $b$ bits over $T_\text{p}$ second and frequency bandwidth $W_\text{p}$. Hence, the PU transmission rate is given by

$$r_{\text{p,pd}} = \frac{b}{W_\text{p} T_\text{p}} \quad (39)$$

When the PU communicates with its destination alone, i.e., without interference from the SU, the link between the PU and the primary destination (i.e., the p $\to$ pd link) is not in outage with probability

$$\overline{\mathbb{P}_{\text{p,pd}}} = \exp\left(-\mathcal{N}_\circ \frac{2^{\frac{b}{W_\text{p} T_\text{p}}} - 1}{P_\circ \sigma_{\text{p,pd}}^2}\right) \quad (40)$$

The probability of primary packet correct decoding at the SU is equal to the probability of the p $\to$ s link being not in outage. This is given by a formula similar to the one in (40) with the relevant parameters of the p → s link. That is,

$$\overline{\mathbb{P}_{\text{p,s}}} = \exp\left(-\mathcal{N}_\text{o}\frac{2^{\frac{b}{W_\text{p}T_\text{p}}}-1}{P_\text{o}\sigma_{\text{p,s}}^2}\right) \quad (41)$$

The SU relays (retransmits) the primary packet over $T_\text{s}$ seconds and frequency bandwidth $W_\text{p}$. Hence, the transmission rate of the relayed primary packet is given by

$$r_{\text{s,pd}} = \frac{b}{W_\text{p}T_\text{s}} \quad (42)$$

The relayed primary packet transmitted by the SU is correctly decoded at the primary destination with probability

$$\overline{\mathbb{P}_{\text{s,pd}}} = \exp\left(-\mathcal{N}_\text{o}\frac{2^{\frac{b}{W_\text{p}T_\text{s}}}-1}{P_\text{o}\sigma_{\text{s,pd}}^2}\right) \quad (43)$$

where (43) is the probability that the s → pd link is not in outage.

## APPENDIX B

When the SU and the PU transmit at the same time over the primary subband, the outage event of the p → pd link is given by

$$\mathbb{P}_{\text{p,pd}}^{(\mathcal{I})} = \Pr\left\{\frac{b}{T_\text{p}W_\text{p}} > \log_2\left(1+\frac{\alpha_{\text{p,pd}}P_\text{o}}{\mathcal{N}_\text{o}+\alpha_{\text{s,pd}}P_\text{o}}\right)\right\} \quad (44)$$

This can be written as

$$\mathbb{P}_{\text{p,pd}}^{(\mathcal{I})} = \Pr\left\{2^{\frac{b}{T_\text{p}W_\text{p}}}-1 > \frac{\alpha_{\text{p,pd}}P_\text{o}}{\mathcal{N}_\text{o}+\alpha_{\text{s,pd}}P_\text{o}}\right\} \quad (45)$$

Since the channels are independent, the region where the inequality $2^{\frac{b}{T_\text{p}W_\text{p}}}-1 > \frac{\alpha_{\text{p,pd}}P_\text{o}}{\mathcal{N}_\text{o}+\alpha_{\text{s,pd}}P_\text{o}}$ is satisfied can be easily obtained. After some algebra, the probability of primary packet correct decoding when the SU interrupts the PU transmission data over $W_\text{p}$ is given by

$$\overline{\mathbb{P}_{\text{p,pd}}^{(\mathcal{I})}} = \frac{\overline{\mathbb{P}_{\text{p,pd}}}}{1+\frac{\sigma_{\text{s,pd}}^2}{\sigma_{\text{p,pd}}^2}(2^{\frac{b}{W_\text{p}T_\text{p}}}-1)} \leq \overline{\mathbb{P}_{\text{p,pd}}} \quad (46)$$

From expression (46), the successful transmission in case of interference is outer bounded by $\overline{\mathbb{P}_{\text{p,pd}}}$. This quantifies the reduction of primary throughput due to concurrent transmission which may occur due to sensing errors. As the message rate increases $\frac{b}{W_\text{p}T_\text{p}}$, the outage probability $\mathbb{P}_{\text{p,pd}}$ increases. Under interference, the correct decoding probability decreases with the same amount in addition to a reduction factor of $(2^{\frac{b}{W_\text{p}T_\text{p}}}-1)$. Moreover, as the cross-channel average gain, given by $\sigma_{\text{s,pd}}^2$, decreases, the correct decoding probability increases since the interference is weak. Actually, what really matters is the ratio of the average of the direct (main) channel and the interference channel, which is given by $\frac{\sigma_{\text{s,pd}}^2}{\sigma_{\text{p,pd}}^2}$. As $\frac{\sigma_{\text{s,pd}}^2}{\sigma_{\text{p,pd}}^2}$ decreases, the interference can cause no impact on the correct packet decoding. When $\frac{\sigma_{\text{s,pd}}^2}{\sigma_{\text{p,pd}}^2}$ or the transmission rate $\frac{b}{W_\text{p}T_\text{p}}$ is very small, we have $\overline{\mathbb{P}_{\text{p,pd}}^{(\mathcal{I})}} \approx \overline{\mathbb{P}_{\text{p,pd}}}$.

## APPENDIX C

In this appendix, we derive the sensing errors probabilities at the SU. The detection problem at time slot $\mathbb{T}$ (assuming that $\tau_\text{s}F_\text{s}$ is an integer, where $F_\text{s}=W_\text{p}$ is the sampling frequency of spectrum sensing [23] and $W_\text{p}$ is the primary bandwidth in case of cooperation) is described as follows

$$\begin{aligned}\mathcal{H}_1: \ & s(\hat{k}) = \zeta_{\text{p,s}}x(\hat{k}) + \varepsilon(\hat{k}) \\ \mathcal{H}_0: \ & s(\hat{k}) = \varepsilon(\hat{k})\end{aligned} \quad (47)$$

$$\mathbb{H}(s) = \frac{1}{F_\text{s}\tau_\text{s}}\sum_{\hat{k}=1}^{F_\text{s}\tau_\text{s}}|s(\hat{k})|^2 \quad (48)$$

where $|\zeta_{\text{p,s}}|^2 = \alpha_{\text{p,s}}$ is channel gain of of the p → s link, hypotheses $\mathcal{H}_1$ and $\mathcal{H}_0$ denote the cases where the PT is active and inactive, respectively, $\tau_\text{s}F_\text{s}$ is the total number of used samples for primary activity detection, $\varepsilon$ is the noise instantaneous value at time slot $\mathbb{T}$ with variance $\mathcal{N}_\text{p} = \mathcal{N}_\text{o}W_\text{p}$, $x$ is the PU transmitted signal at slot $\mathbb{T}$ with variance $P_\text{p} = P_\text{o}W_\text{p}$, $x(\hat{k})$ is the $\hat{k}$-th sample of the PU transmit signal, $s(\hat{k})$ is the $\hat{k}$-th received sample of the primary signal at the SU, and $\mathbb{H}(\cdot)$ is the test statistic of the energy detector.

The quality of the sensing process outcome is determined by the probability of detection, $P_\text{D} = 1 - P_\text{MD}$, and the probability of false alarm, $P_\text{FA}$, which are defined as the probabilities that the spectrum sensing scheme detects a PU under hypotheses $\mathcal{H}_1$ and $\mathcal{H}_0$, respectively. Using the central limit theorem (CLT), the test statistic $\mathbb{H}$ for hypothesis $\mathcal{H}_\theta$, $\theta \in \{0,1\}$, can be approximated by Gaussian distributions [23] with parameters

$$\Lambda_\theta = \theta\alpha_{\text{p,s}}P_\text{p} + \mathcal{N}_\text{p}, \ \sigma_\theta^2 = \frac{(\theta\alpha_{\text{p,s}}P_\text{p}+\mathcal{N}_\text{p})^2}{F_\text{s}\tau_\text{s}} \quad (49)$$

where $\Lambda_\theta$ and $\sigma_\theta^2$ denote the mean and the variance of the Gaussian distribution for the hypothesis $\mathcal{H}_\theta$, where $\theta \in \{0,1\}$. Since $\alpha_{\text{p,s}}$ is Exponentially distributed random variable with parameter $1/\sigma_{\text{p,s}}^2$, the probabilities $P_\text{FA}$ and $P_\text{D}$ can be written as

$$\begin{aligned}P_\text{D} &= \Pr\{\mathbb{H}(s) > \epsilon|\mathcal{H}_1\} \\ &= \frac{\exp(\frac{\mathcal{N}_\text{p}}{\sigma_{\text{p,s}}^2P_\text{p}})}{\sigma_{\text{p,s}}^2P_\text{p}}\int_{\mathcal{N}_\text{p}}^\infty \mathcal{Q}(\sqrt{F_\text{s}\tau_\text{s}}[\frac{\epsilon}{z}-1])\exp(-\frac{z}{\sigma_{\text{p,s}}^2P_\text{p}})dz\end{aligned} \quad (50)$$

$$P_\text{FA} = \Pr\{\mathbb{H}(s) > \epsilon|\mathcal{H}_0\} = \mathcal{Q}(\sqrt{F_\text{s}\tau_\text{s}}[\frac{\epsilon}{\mathcal{N}_\text{p}}-1]) \quad (51)$$

where $\exp(\cdot)$ denotes the exponential function, $\epsilon$ is the energy threshold and $\mathcal{Q}(\mathcal{Y}) = \frac{1}{\sqrt{2\pi}}\int_\mathcal{Y}^\infty \exp(-z^2/2)dz$ is the $\mathcal{Q}$-function.

For a targeted false alarm probability, $P_\text{FA}$, the value of the threshold $\epsilon$ is given by

$$\epsilon = \mathcal{N}_\text{p}\left(\frac{\mathcal{Q}^{-1}(P_\text{FA})}{\sqrt{F_\text{s}\tau_\text{s}}}+1\right) \quad (52)$$

Thus, for a targeted false alarm probability, $P_\text{FA}$, the probability of misdetection is given by substituting Eqn. (52) into Eqn. (50). That is,

$$\begin{aligned}P_\text{MD} = 1 &- \frac{1}{\sigma_{\text{p,s}}^2P_\text{p}}\exp(\frac{1}{\sigma_{\text{p,s}}^2P_\text{p}}\mathcal{N}_\text{p}) \\ &\times \int_{\mathcal{N}_\text{p}}^\infty \mathcal{Q}\left(\sqrt{F_\text{s}\tau_\text{s}}\left(\frac{\mathcal{N}_\text{p}\left(\frac{\mathcal{Q}^{-1}(P_\text{FA})}{\sqrt{F_\text{s}\tau_\text{s}}}+1\right)}{z}-1\right)\right)\exp(-\frac{z}{\sigma_{\text{p,s}}^2P_\text{p}})dz\end{aligned} \quad (53)$$

where $\mathcal{Q}^{-1}(\cdot)$ is the inverse of $\mathcal{Q}$-function.

## APPENDIX D

In this Appendix, we derive the average value of SU instantaneous data rate $R = \log_2(1 + \alpha_{s,sd}\gamma_{s,sd})$. It can be shown that

$$\begin{aligned}\mathcal{G}_s &= \frac{1}{\sigma_{s,sd}^2} \int_0^\infty \log_2(1 + \alpha_{s,sd}\frac{P_\circ}{\mathcal{N}_\circ}) \exp(-\frac{\alpha_{s,sd}}{\sigma_{s,sd}^2}) \ d\alpha_{s,sd} \\ &= \frac{1}{\ln(2)} \exp(\frac{1}{\gamma_{s,sd}\sigma_{s,sd}^2})\Gamma\left(0, \frac{1}{\gamma_{s,sd}\sigma_{s,sd}^2}\right)\end{aligned} \quad (54)$$

where $\Gamma(\cdot, \cdot)$ is the upper incomplete Gamma function.

*Proof.* Let $\gamma_{s,sd} = \frac{P_\circ}{\mathcal{N}_\circ}$. Integration by parts and rearranging the resultant, the expression is given by

$$\begin{aligned}\mathcal{G}_s &= -\int_0^\infty \log_2(1 + \alpha_{s,sd}\gamma_{s,sd}) \ d\exp\left(-\frac{\alpha_{s,sd}}{\sigma_{s,sd}^2}\right) \\ &= \underbrace{-\log_2(1 + \alpha_{s,sd}\gamma_{s,sd}) \exp\left(-\frac{\alpha_{s,sd}}{\sigma_{s,sd}^2}\right)\Big|_0^\infty}_{\text{zero}} \\ &\quad + \frac{1}{\ln(2)} \int_0^\infty \exp\left(-\frac{\alpha_{s,sd}}{\sigma_{s,sd}^2}\right) \frac{\gamma_{s,sd}}{1 + \alpha_{s,sd}\gamma_{s,sd}} \ d\alpha_{s,sd}\end{aligned} \quad (55)$$

After eliminating the zero term, $\mathcal{G}_s$ becomes

$$\mathcal{G}_s = \frac{1}{\ln(2)} \int_0^\infty \exp\left(-\frac{\alpha_{s,sd}}{\sigma_{s,sd}^2}\right) \frac{\gamma_{s,sd}}{1 + \alpha_{s,sd}\gamma_{s,sd}} \ d\alpha_{s,sd} \quad (56)$$

Letting $z = 1 + y$, $\mathcal{G}_s$ becomes

$$\begin{aligned}\mathcal{G}_s &= \frac{1}{\ln(2)} \int_0^\infty \exp\left(-\frac{y}{\gamma_{s,sd}\sigma_{s,sd}^2}\right) \frac{1}{1+y} dy \\ &= \frac{1}{\ln(2)} \int_1^\infty \exp\left(-\frac{z-1}{\gamma_{s,sd}\sigma_{s,sd}^2}\right) \frac{1}{z} dy \\ &= \frac{1}{\ln(2)} \exp\left(\frac{1}{\gamma_{s,sd}\sigma_{s,sd}^2}\right) \int_1^\infty \exp\left(-\frac{z}{\gamma_{s,sd}\sigma_{s,sd}^2}\right) \frac{1}{z} dz\end{aligned} \quad (57)$$

Letting $\frac{z}{\gamma_{s,sd}\sigma_{s,sd}^2} = q$, we get

$$\begin{aligned}\mathcal{G}_s &= \frac{1}{\ln(2)} \exp\left(\frac{1}{\gamma_{s,sd}\sigma_{s,sd}^2}\right) \int_{\frac{1}{\gamma_{s,sd}\sigma_{s,sd}^2}}^\infty \exp(-q) \frac{1}{q} dq \\ &= \frac{1}{\ln(2)} \exp\left(\frac{1}{\gamma_{s,sd}\sigma_{s,sd}^2}\right) \Gamma\left(0, \frac{1}{\gamma_{s,sd}\sigma_{s,sd}^2}\right)\end{aligned} \quad (58)$$

where $\Gamma(\cdot, \cdot)$ is the upper incomplete Gamma function. □